# MSDiff: Multi-Scale Diffusion Model for Ultra-Sparse View CT Reconstruction

Junyan Zhang, Mengxiao Geng, Pinhuang Tan, Yi Liu, Zhili Liu, Bin Huang, Qiegen Liu, *Senior Member, IEEE*

*Abstract*—**Computed Tomography (CT) technology reduces radiation exposure to the human body through sparse sampling, but fewer sampling angles pose challenges for image reconstruction. When the projection angles are significantly reduced, the quality of image reconstruction deteriorates. To improve the quality of image reconstruction under sparse angles, an ultra-sparse view CT reconstruction method utilizing multi-scale diffusion models is proposed. This method aims to focus on the global distribution of information while facilitating the reconstruction of local image features in sparse views. Specifically, the proposed model ingeniously combines information from both comprehensive sampling and selective sparse sampling techniques. By precisely adjusting the diffusion model, diverse noise distributions are extracted, enhancing the understanding of the overall image structure and assisting the fully sampled model in recovering image information more effectively. By leveraging the inherent correlations within the projection data, an equidistant mask is designed according to the principles of CT imaging, allowing the model to focus attention more efficiently. Experimental results demonstrate that the multi-scale model approach significantly improves image reconstruction quality under ultra-sparse views and exhibits good generalization across multiple datasets.**

## I. Introduction

Computed tomography (CT) utilizes X-ray to achieve tomographic imaging of the human body, providing crucial three-dimensional internal structural information. It plays an indispensable role in clinical diagnosis [1], [2]. However, statistics show that up to 60% of CT examinations use an excessively high radiation dose [3], which may significantly increase patient cancer risks [4]. Currently, one of the main methods to reduce radiation dosage is to decrease the number of projection angles [5]. At present, one of the primary strategies for radiation dose reduction in CT imaging is to decrease the number of projection views, commonly referred to as sparse-view acquisition. In addition to this, dose can also be reduced by lowering the initial X-ray beam intensity or photon counts. While both approaches effectively decrease patient exposure, they introduce distinct reconstruction challenges—sparse-view acquisition leads to incomplete data, whereas low-intensity acquisition increases noise levels in the measured projections. There are generally two strategies to reduce projection angles: One is to conduct projections over the 360° range but increase the interval between projection angles, known as sparse-angle reconstruction, the other is continuous projection over a range less than 180°, referred to as limited-angle reconstruction. Sparse-angle reconstruction is more frequently employed with the explicit aim of dose reduction, whereas limited-angle acquisition tends to be used in situations where full angular sampling is infeasible. To effectively reduce CT radiation dosage, researchers have explored various algorithms to reconstruct full-view CT images from sparse-view data. Chen *et al.* [6] proposed a limited-angle CT reconstruction method based on anisotropic TV minimization, which can reconstruct full-view CT images from sparse-angle CT data by leveraging prior knowledge and mathematical optimization. In the field of deep learning, Lee *et al*. [7] synthesized complete sinusoidal graphs with deep neural net-works to reconstruct full-view CT. Zhang *et al*. [8] built a DREAM-Net, leveraging projection completion, residual learning and other modules collaboratively for full-view CT reconstruction. However, the commonly used sparse-view CT acquisition pattern today typically requires scanning and collecting projection data from 60-180 view angles [9], [10]. Although radiation dose is reduced by 50%-80% compared to full-view CT scanning, such levels remain too high for certain vulnerable groups with potential hazards [11], [12]. Moreover, scanning 60-180 views takes time, unfriendly to emergency patients needing rapid diagnosis [13]. Consequently, a reduction of CT views to 60 or fewer leads to a significant decrease in both radiation exposure and scanning time for patients. This approach renders the technology more suitable for vulnerable groups and emergency cases.

Ultra-sparse view CT is a sampling mode in computed tomography (CT) that uses a very small number of projection angles, usually much lower than the number of angles required for conventional CT scans, typically using only 5%–10% of the total viewing angle, that is, once every 10 degrees or even once every 20 degrees. Adopting ultra-sparse view can tremendously reduce CT radiation dose, better achieving the ALARA principle and alleviating radiation harm to patients [14]. However, ultra-sparse view leads to severe stripe pseudo-artifacts in the reconstructed images. In this case, the system matrix is severely underdetermined, and

This work was supported in part by National Natural Science Foundation of China under 62122033 and Key Research and Development Program of Jiangxi Province under 20212BBE53001. (J. Zhang and M. Geng are co-first authors.) (Corresponding authors: B. Huang and Q. Liu)

J. Zhang, P. Tan, M. Geng, Y. Liu and Q. Liu are with the School of Information Engineering, Nanchang University, Nanchang 330031, China. ({zhangjunyan, tanpinhuang, mxiaogeng, liuyi}@email.ncu.edu.cn, liuqiegen@ncu.edu.cn)

B. Huang are with School of Mathematics and Computer Sciences, Nanchang University, Nanchang 330031, China. ({huangbin}@email.ncu.edu.cn)

Z. Liu are with Department of Orthopedic Surgery, The First Hospital of Nanchang, The Third Affiliated Hospital of Nanchang University, Nanchang, China. (zhili-liu@ncu.edu.cn)

ultra-sparse angular CT usually relies on reconstruction methods based on sparse representation, compressed sensing (CS), deep learning, or model-driven deep networks to recover high-quality tomographic images from very little angle data. Wu *et al.* proposed a data-iterative optimization score-based model (DOSM), incorporating data consistency into SDE for ultra-sparse-view CT reconstruction, achieving excellent performance with minimal views [15]. Chan *et al*. proposed an end-to-end attention-based network, first reconstructing with filtered back projection (FBP) and then removing artifacts [16]. Although some progress has been made, the image quality still hardly meets diagnostic needs. How to obtain acceptable diagnostic CT images under ultra-low radiation dose remains a significant challenge. Recently, Guan *et al*. adopted a fully unsupervised score-based generative model for sparse-view CT reconstruction, showing it can achieve quality comparable to supervised adversarial learning [17]. Additionally, Xu *et al*. applied diffusion models to progressively reconstruct sparse-view CT in the wavelet domain, also achieving remarkable improvement in image quality [18]. Wang *et al*. [19] proposed a sub-volume-based 3D denoising diffusion probabilistic model (DDPM) for CBCT image reconstruction from sparsely-sampled data. Xia *et al*. [20] presented a patch-based DDPM for sparse-view X-ray computed tomography reconstruction. Li *et al*. [21] introduced a framework named DDMM-Synth combining diffusion models with CT measurement embedding inverse sampling schemes for sparse CT reconstruction. Huang *et al*. [22] proposed a projection-domain fully unsupervised single-sample diffusion model (OSDM) for low-dose CT reconstruction, which takes abundant tensors extracted from the structure-Hankel matrix as network input to capture prior distribution. These further prove the efficacy of diffusion models in suppressing noise and restoring details by controlling random processes, enabling high-quality reconstruction for sparse-view CT.

Despite the relative success achieved by existing diffusion models in sparse CT image reconstruction, ultra-sparse CT reconstruction poses challenges due to the lack of sufficient projection data. To address this issue, this study proposes a multi-scale diffusion model for ultra-sparse view CT reconstruction. This method involves alternating iterative reconstruction using multiple diffusion models with diverse prior distribution, aiming to significantly enhance the quality of CT images reconstructed from highly sparse views. In summary, the main contributions of this work can be summarized as follows:

- We propose a reconstruction method for a multi-scale diffusion model to enhance the restoration of local textures and details. Due to the extremely limited information contained in the ultra-sparse views, our method aims to fully utilize existing data and integrate the extraction of local projection information by the model to compensate for the challenges caused by data missing. Integrating local projection with structural details enhances both the accuracy of image reconstruction and the restoration of local details, while ensuring global consistency.
- We employ the existing correlation in projection data to design specialized masks for extracting sparse priors. This method utilizes fewer prior cues, thereby enhancing the comprehension of the model for sparse data distribution. At the same time, the model can more precisely capture crucial features within the data, thus facilitating a more accurate restoration of local textures and details during the reconstruction process.

The structure of this work is arranged as follows: Section II provides a brief introduction to the relevant knowledge of CT reconstruction. Section III elaborates on the theory and algorithms of CT ultra-sparse projection reconstruction based on multiple diffusion models. Section IV showcases the experimental results and evaluations under ultra-sparse angle conditions. Finally, this work is summarized in Section V.

## II. Preparation

### A. CT Reconstruction Model

CT image reconstruction starts with acquiring projection data, mathematically represented by the forward projection equation:

$$x = AI, \tag{1}$$

where $x$ represents the vector of measured projection data, and $A$ symbolizes the system matrix, which encompasses the physics of the CT imaging process. The system matrix $A$ models the line integrals through the attenuation coefficients that are associated with the subject, which are discretely represented by the vector $I$. Thus, the product $AI$ effectively simulates the projection data as if derived from the actual object.

To reconstruct images from projections, the optimization problem is solved as follows:

$$\min_I \{\|AI - x\|_2^2 + \lambda R(I)\}, \tag{2}$$

where $\|AI - x\|_2^2$ is the data fidelity term and $R(I)$ is the regularization term to describe prior knowledge of the image. $\lambda$ is a positive parameter used to balance these terms.

The transition from full projection data to sparse-view data can be interpreted as a linear measurement process. Fig. 1 provides an intuitive visualization of this linear measurement process. By applying the Radon transform $r$, the original image is transformed into the projection data. Additionally, $m$ represents the subsampling mask for the projection data, and $P(m)$ subsamples the sinogram to sparse-view data with subsampling mask. Consequently, the sparse-view CT reconstruction problem can be formulated as:

$$y = P(m)AI = P(m)x, \tag{3}$$

where $y$ is the sparse-view CT projection data and $x$ represents the full-view projection data.

In scenarios involving sparse-view projection, strip artifacts are commonly produced when images are reconstructed directly using the FBP algorithm. To obtain higher-quality reconstructed images from sparse-view projection data, two general strategies are commonly adopted. One is to complete or synthesize the missing projections in the sinogram domain, thereby approximating full-view data before reconstruction. The other is to directly reconstruct the image from sparse-view data and subsequently perform artifact suppression or correction in the image domain, a process that constitutes an underdetermined inverse problem. The former approach explicitly addresses the data-domain incompleteness but may propagate interpolation errors, whereas the latter avoids sinogram synthesis but relies on strong image priors or deep learning models to mitigate streak artifacts. To address this problem, various priors are incorporated into a regularized objective function, which is expressed as follows:

$$x^* = \underset{x}{\operatorname{argmin}}\{\|P(m)x - y\|_2^2 + \lambda R(x)\}. \tag{4}$$

The objective function consists of two terms. The first term represents the data fidelity, ensuring that the sparse-view sinogram aligns with the estimated measurements obtained through the subsampling mask. The second term is the regularization term, where $\lambda$ helps maintain an appropriate balance between data fidelity and regularization term.

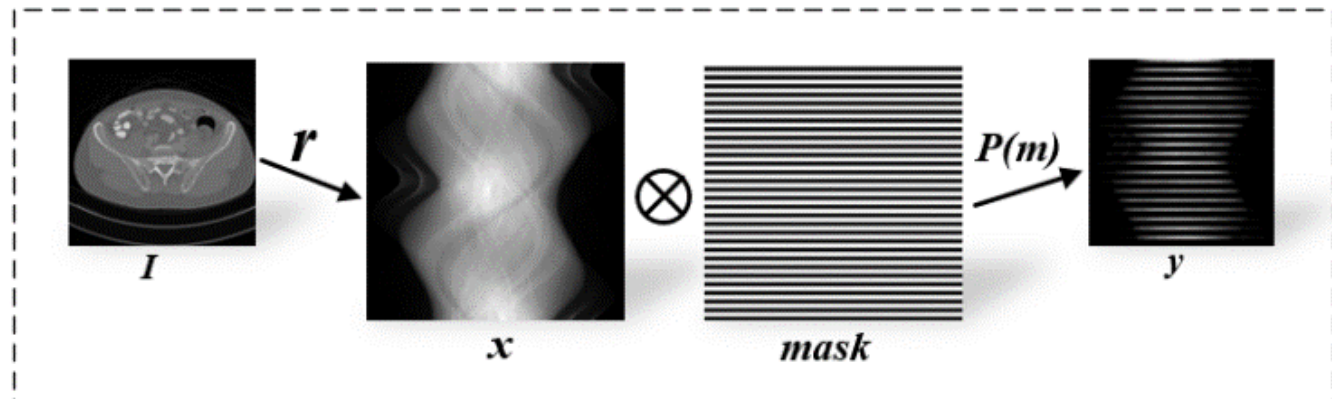

**Fig. 1.** Linear measurement process for sparse-view CT.

## B. Diffusion Model

In recent years, deep generative models based on diffusion processes have shown great potential in modeling complex high-dimensional data distribution [23], [24]. Diffusion-based models can be categorized into two main types:

***Denoising Diffusion Probabilistic Models (DDPMs):*** Sohl-Dickstein *et al.* first proposed learning the reverse diffusion process to infer data distribution [25], which inspired DDPMs. A typical example is DDPMs proposed by Ho *et al.* [26], which learns distribution by simulating the reverse diffusion of data generation processes. DDPMs use Gaussian noise and assumes a Markov diffusion process. Researchers have made various improvements to DDPMs, including using non-Markovian processes [27], mixed Gaussian noise [28], forming models like the denoising diffusion implicit model (DDIM). These DDPMs demonstrate strong ability in detail recovery and noise control.

***Score-based Models with Stochastic Differential Equations (SDEs):*** Song *et al.* [29] introduced SDEs with forward and reverse-time SDEs into diffusion models. SDEs provide a mathematical framework to derive the score function. Models based on SDEs like Song *et al.* [30] and Wang *et al.* [31] have shown promising results. Compared to networks such as generative adversarial networks (GANs), diffusion-based models can better capture complex data structures [32]. They pioneer a new paradigm for learning complex distribution.

The diffusion model encompasses both forward and reverse stochastic differential equations. The forward stochastic differential equation can be represented by the following formula:

$$dx = f(x,t)dt + g(t)d\boldsymbol{w}, \tag{5}$$

where $f(x,t) \in \mathbb{R}^n$ represents the drift coefficient, and $g(t)$ denotes the diffusion coefficient. The term $\boldsymbol{w} \in \mathbb{R}^n$ introduces Brownian motion. To be specific, the variance exploding VE-SDE configuration $(f(x,t) = 0, g(t) = \sqrt{d[\sigma^2(t)]/dt})$ is embraced, a choice that imparts heightened generative prowess. Based on this, assigning the VE-SDE configuration to Equation 5, the forward process of VE-SDE can be expressed as:

$$\mathrm{d}x = \sqrt{d[\sigma^2(t)]/dt}\,\mathrm{d}\boldsymbol{w}, \tag{6}$$

where $\sigma(t) > 0$ is represented as a monotonically increasing function. It is thoughtfully chosen to manifest as a geometric series. Samples of $x(T) \sim P_T$ are started from, and through the reversal of the process, samples $x(0) \sim P_0$ are obtained. It is a recognized principle in stochastic calculus that the reverse process of diffusion retains the characteristics of diffusion. Thus, the following reverse-time SDE is obtained:

$$\begin{aligned} dx &= [f(x,t) - g^2(t)\nabla_x \log p_t(x)]dt + g(t)d\boldsymbol{w} \\ &= \frac{\mathrm{d}[\sigma^2(t)]}{dt}\nabla_x \log p_t(x) + \sqrt{\frac{\mathrm{d}[\sigma^2(t)]}{dt}}\,d\boldsymbol{w}, \end{aligned} \tag{7}$$

It is assumed that $x(0)$ follows the initial data distribution of the projection views $p_0$, and $x(T)$ conforms to the prior distribution $p_t$. To substitute for the unknown $\nabla_x \log p_t(x)$, a denoising score matching technique is employed, where $\nabla_x \log p_t((x)|x(0))$ represents data centered around a Gaussian perturbation kernel at $x(0)$. Under certain regularization conditions, as specified in (7), the score function $S_\theta(x,t)$ trained with denoising score matching is nearly certain to satisfy the target function $S_\theta(x,t) \approx \nabla_x \log p_t(x)$. Through systematic training of the network $S_\theta$, the calculation of score functions is approximated, thereby capturing the essence of $\nabla_x \log p_t(x)$, which is represented as $S_\theta(x,t) = \nabla_x \log p_t(x(t))$. By training the score network to obtain the data distribution $S_\theta(x,t)$, it is inserted into (7) and the resulting reverse-time SDE is solved for reconstruction.

$$dx = [f(x,t) - g^2(t)S_\theta(x,t)]dt + g(t)d\boldsymbol{w}. \tag{8}$$

Furthermore, it is necessary to fine-tune the neural network's parameters $\theta^*$, to accurately estimate $\nabla_x \log p_t(x)$. This process is considered the core objective function of a score-based Stochastic Differential Equation (SDE):

$$\theta^* = \underset{\theta}{\operatorname{argmin}} \mathbb{E}_t\{\lambda(t)\mathbb{E}_{x(0)}\mathbb{E}_{x(t)|x(0)}[||S_\theta(x(t),t) - \nabla_{x(t)} \log p_t\,(x(t)|x(0))||_2^2]\}, \tag{10}$$

where $\lambda(t)$ represents a positively weighted function, and $t$ is uniformly sampled from the interval $[0, T-1]$. The term $p_t(x(t)|x(0))$ signifies the Gaussian perturbation kernel centered around $x(0)$. A neural network $S_\theta$ is systematically trained to approximate the score, allowing it to capture the essence of $\nabla_x \log p_t\,(x)$, *i.e.*, $S_\theta = \nabla_x \log p_t\,(x)$. Upon achieving full model proficiency, the value of $\nabla_x \log p_t\,(x)$ can be effectively determined across all time points $t$ through the solution of $S_\theta(x(t),t)$. This equips the model with an ample reservoir of prior knowledge, thereby enabling it to skillfully undertake the process of image reconstruction.

Building upon the trained score model, the Predictor–Corrector (PC) sampler is typically employed to iteratively generate high-quality reconstructions. The predictor step integrates the learned stochastic dynamics derived from the reverse SDE to propagate the sample toward the data manifold, while the corrector step refines this trajectory through additional Langevin-type perturbations, effectively reducing sampling bias and stabilizing convergence. This two-stage mechanism enables a balance between stochastic exploration and deterministic refinement, ensuring that the reconstructed images not only adhere to the learned data distribution but also preserve fine structural details essential for high-fidelity CT restoration.

## III. Method

### A. Motivation

CT projection data are typically represented in graphs, containing complex features and structural information of projection images. Sparse projection data, acquired from limited angles, are designated as such. Compared to complete projection data, sparse projection data are found to be more sensitive to noise. While traditional reconstruction algorithms can be effective under certain conditions, a diminishment in performance is observed with reduced projection angles, especially in ultra-sparse views. Streaking and aliasing artifacts may be produced during the reconstruction of ultra-sparse projection data, posing significant challenges in ultra-sparse view scenarios. In this study, a novel model for reconstructing ultra-sparse views is proposed to enhance the overall visual quality of images. Specifically, two diffusion models, the full-view diffusion model (FDM) and the sparse-view diffusion model (SDM), are trained to extract global and specific local information from fully sampled and sparsely projected data, respectively. More precisely, the overall characteristics of the projection data are effectively captured by FDM, which also reduces noise and artifacts. The reconstruction process is assisted by SDM, which complements FDM by highlighting intrinsic connections between structures and demonstrating unique advantages in structures, edges, textures, and other aspects. By integrating multiple diffusion models to mutually constrain each other, the stability of ultra-sparse view CT reconstruction is significantly enhanced. The reconstruction of ultra-sparse view is achieved by iteratively processing specific sparse projection data, as illustrated in Fig. 2. From left to right, sparse projection data are sequentially reconstructed to approximate the ground truth (GT).

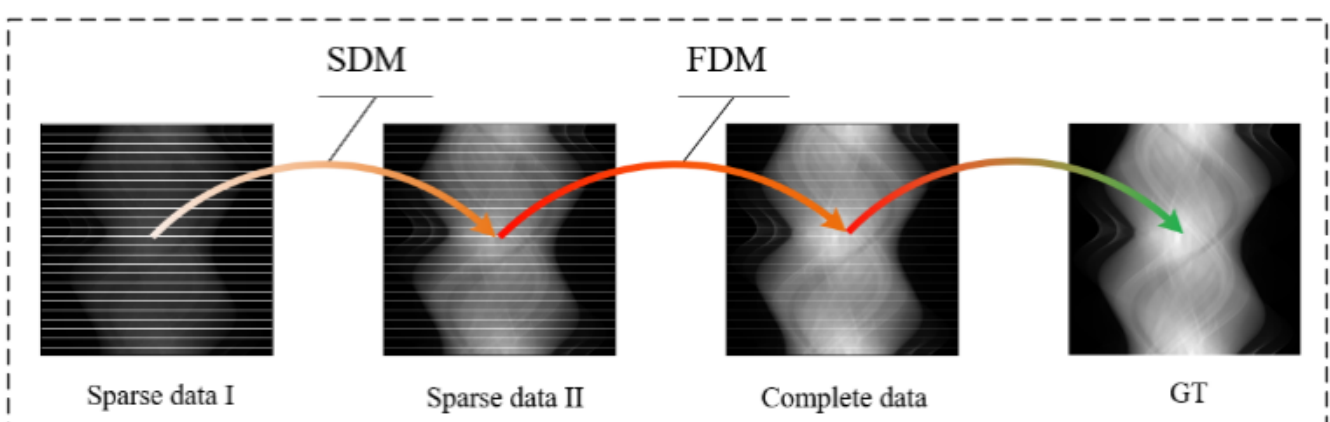

**Fig. 2.** The proposed reconstruction scheme integrates both global and local information. SDM represents targeted reconstruction for specific subsets of data. FDM represents global information reconstruction for various image features.

### B. Prior Learning in Projection Domain

During the model training process, SDM and FDM are trained separately using masked sparsely-sampled sinogram data along with completely fully-sampled data. To simplify and accelerate the explanation, the shorthand notation $x_1$ is introduced for full-view projection data, and $x_2$ for sparse-view projection data. The specific masking operator is defined as follows:

$$x_1 = x, \tag{11}$$

$$x_2 = x \odot mask, \tag{12}$$

where $mask$ is the specific mask matrix operator used to separate the information for training. The horizontally striped $mask$ is applied to the projection-domain image $x \in \mathbb{R}^{v \times d}$, where $v$ and $d$ denote the numbers of projection views and detector elements, respectively. The mask selectively removes angular information, generating $x_2$ as a sparse-view counterpart of $x$. This formulation naturally supports a two-stage modeling paradigm that jointly captures the characteristics of sparse-view reconstruction and high-fidelity reconstruction. $x$ represents input fully-sampled projection data. $\odot$ represents element-wise multiplication. At specific sparse sampling angles, row matrices assigned for partial information extraction are assigned a value of 1, whereas other areas are marked with a value of 0. Therefore, areas falling outside the designated sparse sampling angles are ignored, allowing the algorithm to focus solely on data captured at positions marked with 1. These positions, identified by row matrices valued at 1, are recognized as sampling points under consideration, facilitating the preservation of potential structural information within the image. Conversely, areas marked with a value of 0 are disregarded, which aids in reducing noise introduction. This approach effectively reduces the volume of data and assists the model in concentrating on learning and reconstructing fundamental data characteristics.

The utilization of the SDE model to learn the statistical distribution characteristics of CT sinogram data is illustrated in Fig. 3. Fig. 3(a) illustrates the diffusion process in the projection domain, where both FDM and SDM follow the same process. Sparse projection data from 120 views are used to train SDM in Fig. 3(b). $S_{\theta_1}$ and $S_{\theta_2}$ respectively denote the training score networks associated with FDM and SDM. Specifically, Gaussian noise is continuously added to the images, and through the encoder and decoder, a transition is gradually made from a coarse prior distribution to an accurate posterior distribution, thereby enabling the generation of clear reconstructed images.

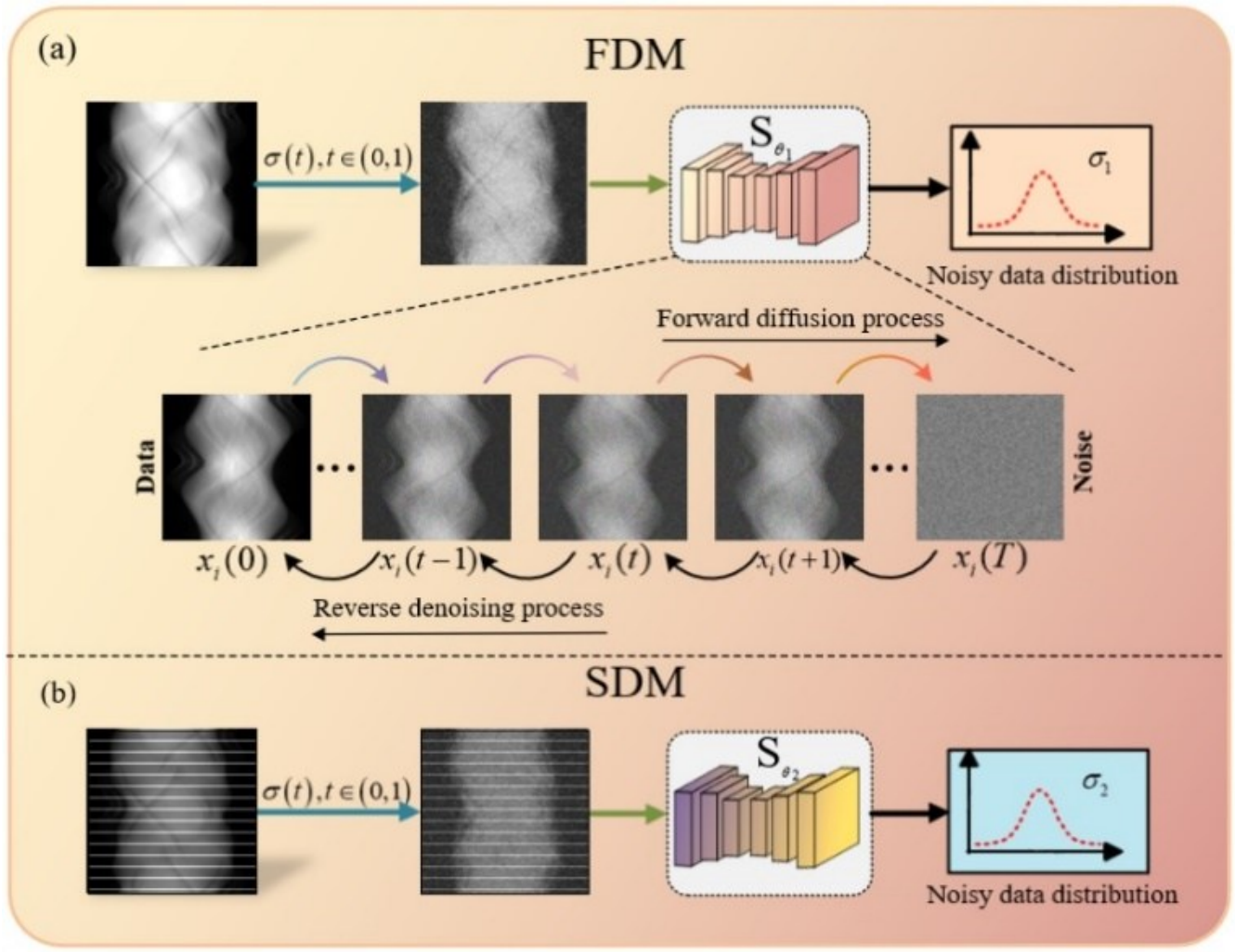

**Fig. 3.** Visualization of the diffusion process of MSDiff in the projection domain. From top to bottom: (a) Detailed training procedure of FDM, including the forward diffusion and reverse diffusion processes. (b) Detailed training procedure of SDM.

The original fully-sampled sinogram data and sparsely-sampled sinogram data are used as inputs to simulate the forward diffusion process of the stochastic differential equation (SDE) [27], allowing the model to gradually approximate the actual distribution of different sinogram data. Base on this, we will introduce the modeling process of Full-view Diffusion Model (FDM) and Sparse-view Diffusion Model (SDM) respectively.

***Full-view Diffusion Model (FDM):*** The study proposes to use FDM as a model for CT image reconstruction. Specifically, in the training stage, the projection domain data of the full-view CT $x_1$ , whose formulation is defined in Eq. (11), is used as prior information, and the configuration of VE-SDE is adopted. Upon the incorporation of $x_1$ into (5), we can get the forward process of FDM as:

$$\mathrm{d}x_1 = \sqrt{d[\sigma_1^2(t)]/dt}\,\mathrm{d}\boldsymbol{w}, \tag{13}$$

The objective function for this optimization is:

$$\theta_1^* = \underset{\boldsymbol{\theta}}{\operatorname{argmin}} \mathbb{E}_t\{\lambda(t)\mathbb{E}_{x_1(0)}\mathbb{E}_{x_1(t)|x_1(0)}[||S_{\theta_1}(x_1(t),t) \\ -\nabla_{x_1(t)} \log p_t\,(x_1(t)|x_1(0))||_2^2]\}, \tag{14}$$

where $\theta_1^*$ represents the optimization objective of the full-views diffusion model.

***Sparse-view Diffusion Model (SDM):*** In order to make up for the difficulty of FDM in adapting to ultra-sparse view data reconstruction, this paper introduces Sparse-view Diffusion Model (SDM) into the reconstruction framework. SDM takes the CT projection domain data of sparse view $x_2$ ,whose formulation is provided in Eq. (12), as the prior-input information, which is fed into a score-based network to learn the sparse-view data distribution. In the sparse-view diffusion model, the forward process of the VE-SDE is expressed as:

$$\mathrm{d}x_2 = \sqrt{d[\sigma_2^2(t)]/dt}\,\mathrm{d}\boldsymbol{w}, \tag{15}$$

The objective function for this optimization is:

$$\theta_2^* = \underset{\boldsymbol{\theta}}{\operatorname{argmin}} \mathbb{E}_t\{\lambda(t)\mathbb{E}_{x_2(0)}\mathbb{E}_{x_2(t)|x_2(0)}[||S_{\theta_2}(x_2(t),t) \\ -\nabla_{x_2(t)} \log p_t\,(x_2(t)|x_2(0))||_2^2]\}, \tag{16}$$

where $\theta_2^*$ represents the optimization objective of the sparse-view diffusion model.

The parameters in above formulas follow the definitions in Eqs. (5) – (10). Learning the distribution of specific sparse projection data enables the extraction of relevant structural information. Coupled with the comprehensive generation capability of the full-view projection model, this enhances the adaptability and generalization ability of the model to data with different sampling rates. Therefore, the low-sampling-rate training strategy improves the modeling and reconstruction ability of the model for sparse-view data to some extent, which is an effective method to enhance the reconstruction quality. **Algorithm 1** outlines the training process: training both the full-view projection model and the specific sparse-view projection model to capture prior distribution.

**Algorithm 1: MSDiff for Training Process**

Generating projection $x$
Under sampling: $x_1 = x,\ x_2 = x \odot mask$
Training datasets distribution: $x_1$ *and* $x_2$
Training with (14) and (16)
Output: $S_{\theta_1}(x_1, t)$ and $S_{\theta_2}(x_2, t)$

### *C. MSDiff: Iterative Reconstruction*

This section describes the iterative reconstruction process of MSDiff. Fig. 4 intuitively displays a two-stage iterative method that combines FDM and SDM for image reconstruction. Initially, the ultra-sparse view sinogram $U$ obtained from the CT device is interpolated using bilinear interpolation to achieve a preliminarily restored full-view sinogram $x$.

$$x = K(U), \tag{17}$$

where $K$ represents a bilinear interpolation operation. Initial reconstruction for ultra-sparse projection data is conducted using SDM, with the sampling scheme determined by the sparse prior of SDM. Consequently, the full-view sinogram is element-wise multiplied by the corresponding $mask$ to produce sparse-view images for SDM reconstruction. The overall framework of the iterative reconstruction process is described as follows:

$$\begin{cases} x'_{i-1,j} = SDM(x_{i,j} \odot mask), \\ x_{i-1,j} = x'_{i-1,j} + (1 - mask) \odot x_{i,j}, \\ x_{i-1,j-1} = FDM(x_{i-1,j}), \end{cases} \tag{18}$$

where $i$ and $j$ denote the reverse iterations for SDM and FDM, respectively. $x_{i-1,j-1}$ and $x'_{i-1,j}$ represent the output results of networks utilizing different prior information corresponding to FDM and SDM, respectively. Since SDM has been trained on the characteristic distribution of specific data, a limited yet precise amount of prior knowledge is provided to guide the reconstruction process, thereby effectively restoring the structural information and texture features within the sinogram data.

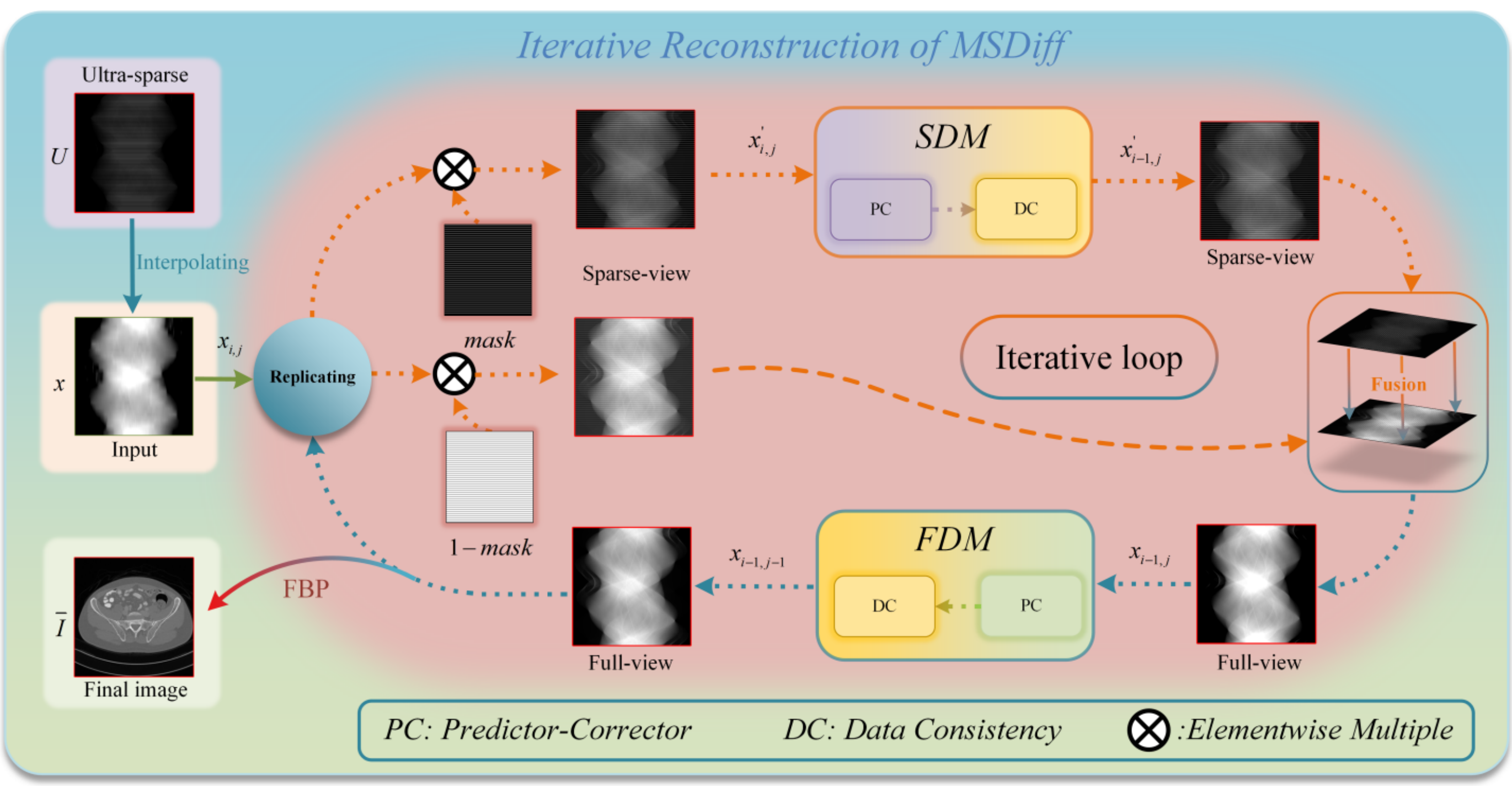

**Fig. 4.** Pipeline of the ultra-sparse view reconstruction procedure in MSDiff. Ultra-sparse view reconstruction is conducted in an iterative manner.

Subsequently, the reconstructed sparse views are superimposed with the sparse views reconstructed by SDM, resulting in the full-view sinogram $x_{i-1,j}$. Thereafter, the sinograms aligned through iterative reconstruction using FDM gradually restore richer details until convergence is achieved, yielding sinograms that closely approximate the original input. Upon completion of the MSDiff iterations, by integrating processing results from multiple scale levels, the MSDiff model produces the final reconstructed result. Eventually, the reconstructed image is obtained using the FBP algorithm. The expression is described as follows:

$$\bar{I} = FBP(x), \tag{19}$$

where $\bar{I}$ stands as the final reconstruction results produced by MSDiff method. It is noteworthy that the predictor-corrector (PC) sampler step is utilized in the model. This work introduces PC sampling at the samples updating step, as suggested in [33]. The predictor is viewed as a numerical solver for the reverse-time SDE in the PC sampling. Once the reverse-time SDE process concludes, samples are generated based on the discretized prior distribution, which can be discretized as follows:

$$x_k \leftarrow x_{k+1} + (\sigma_{k+1}^2 - \sigma_k^2) S_\theta(x_{k+1}, \sigma_{k+1}) + \sqrt{\sigma_{k+1}^2 - \sigma_k^2}\, z$$
$$k = T-1, \cdots, 0, \tag{20}$$

where $z \sim \mathcal{N}(0,1)$ refers to a standard normal distribution, $x(0) \sim p_0$, and $\sigma_0 = 0$ is chosen to simplify the notation. The above formulation is repeated for $k = T-1, \cdots, 0$. During the reconstruction process of MSDiff, $k$ is represented by either $i$ or $j$. When it comes to the corrector, it refers to the Langevin dynamics via transforming any initial sample $x(t)$ to the final sample $x(0)$ with the following procedure:

$$x_{k,l} = x_{k,l-1} + \varepsilon_k S_\theta(x_{k,l-1}, \sigma_k) + \sqrt{2\varepsilon_k}\, z$$
$$l = 1,2, \cdots, M, k = T-1, \cdots, 0. \tag{21}$$

where $\varepsilon_k > 0$ is the step size, and the above equation is repeated for $l = 1,2, \cdots, M,\ k = T-1, \cdots, 0$. The theory of Langevin dynamics guarantees that when $M \to 0$ and $\varepsilon_k \to 0$ , $x_k$ is sampled from $p_t(x)$ under designated conditions.

Moreover, to ensure that the output remains consistent with the original data, data consistency (DC) processing is performed on the reconstruction results after each step of the PC sampler. According to (4), the minimization problem is succinctly described as follows:

$$\min_x \{\|P(m)x - y\|_2^2 + \lambda \|x - x_\tau\|_2^2\}, \tag{22}$$

where $x_\tau$ represents the sinogram projection data generated by the network, while $x$ also denotes the sinogram projection data to be reconstructed. Therefore, the corresponding solutions can be derived through the application of first-order optimality conditions:

$$x^*(v) = \begin{cases} x_\tau(v), & if\ v \notin \Omega \\ \dfrac{[P^T(m)y(v) + \lambda x_\tau(v)]}{1+\lambda}, & if\ v \in \Omega \end{cases} \tag{23}$$

where $\Omega$ is represented by specific locations sampled within the projection domain. $x_\tau(v)$ represents the entry at index $v$ in projection domain generated by the network. Furthermore, a detailed operational guide for the training and inference stages is provided by **Algorithm 2**, ensuring the efficiency and accuracy of the entire reconstruction process.

**Algorithm 2: MSDiff for Iterative Reconstruction Process**

**1: Setting:** $S_{\theta_1}, S_{\theta_2}, T, \sigma, \varepsilon$
**2: Initial data:** $x = K(U)$
**3:** $\{x^T\} \sim \mathcal{N}(0, \sigma_{max}^2 \mathbf{I})$
**4:** **For** $i, j = T-1$ to 0 **do**
**5:** $x'_{i,j} = x_{i,j} \odot mask$ , $x^*_{i,j} = (1 - mask) \odot x_{i,j}$
**6:** Update $x'_{i-1,j} \leftarrow Predictor(x'_{i,j}, \sigma_i, \sigma_{i+1}, S_{\theta_2})$ via (20)
**7:** Update $x'_{i-1,j}$ via (23)
**8:** Update $x'_{i-1,j} \leftarrow Corrector(x'_{i-1,j}, \sigma_i, \varepsilon_i, S_{\theta_2})$ via (21)
**9:** Update $x'_{i-1,j}$ via (23)
**10:** $x_{i-1,j} = x'_{i-1,j} + x^*_{i,j}$
**11:** Update $x_{i-1,j-1} \leftarrow Predictor(x_{i-1,j}, \sigma_i, \sigma_{i+1}, S_{\theta_1})$ via (20)
**12:** Update $x_{i-1,j-1}$ via (23)
**13:** Update $x_{i-1,j-1} \leftarrow Corrector(x_{i-1,j-1}, \sigma_i, \varepsilon_i, S_{\theta_1})$ via (21)
**14:** Update $x_{i-1,j-1}$ via (23)
**15:** **End for**
**16:** Obtain the final result by (19)
**17: Return** $\bar{I}$

In summary, the MSDiff processes images by considering various levels of detail and structure, from coarse to fine, simulating a transition from large-scale (global features) to small-scale (local details). The model analyzes and processes images across multiple dimensional layers, enhancing local features while maintaining global consistency. To validate the effectiveness of the proposed multi-scale diffusion model reconstruction strategy, a single FDM is also constructed as a comparative method. This baseline model is trained solely on complete sinogram data, with its reconstruction process including only a single iteration using FDM. Specific quantitative evaluation metrics and comparison results are provided in the experimental section.

## IV. Experiments

### A. Data Specification

***AAPM Challenge Data:*** In this study, simulated abdominal imaging data from the 2016 AAPM CT Low Dose Grand Challenge [34], provided by Mayo Clinic, are initially utilized. A total of 5338 normal-dose CT images with a resolution of 512×512 and a thickness of 1mm are selected and divided into training, validation, and testing sets, with 4742 slices designated for training, 584 for validation, and 12 for testing. Projection data are simulated by adding Poisson noise to the sinograms generated from the normal-dose images to create artifact-free images. Artifact-free images are obtained using the FBP algorithm with 720 projection angles, serving as the standard reference. Subsets of 30, 48, and 60 views are extracted from the full sinogram to evaluate the multi-scale diffusion model algorithm for ultra-sparse view CT reconstruction.

***CIRS Phantom Data:*** A high-quality CT volume dataset measuring 512×512×100 voxels, with each voxel size at 0.78×0.78×0.625, is acquired exclusively. This dataset is obtained using a humanoid CIRS phantom on the GE Discovery HD750 CT system. The distance of source-to-axis is recorded as 573, and the distance of source-to-detector as 1010. Full projection data from 720 views are collected, from which subsets of 10, 20, and 30 views are extracted for algorithm performance evaluation. To further validate the robustness of the proposed method, prior knowledge acquired from the AAPM Challenge data is used to assess its performance on the CIRS Phantom dataset.

***Preclinical Mouse Data:*** To validate the effectiveness of the proposed MSDiff in preclinical micro-CT applications, the networks trained on AAPM challenge data within MSDiff are transferred to preclinical mouse data. Deceased mouse is scanned using an industrial CT system equipped with a microfocus X-ray source and a flat-panel X-ray detector. The distance from the source to the detector and the object are 1150 and 950, respectively. The detector consists of 1024×1024 pixels, each detector unit measuring 0.2×0.2. Projections consisting of 500 views are distributed within the angular range. The reconstructed image forms a matrix of size 512×512, with each pixel measuring 0.15×0.15. With 500 views distributed within the angular range, 50 views are obtained across the scanning range when the subsampling factor is set to 10.

### *B. Model Training and Parameter Selection*

In the experiments, FDM and SDM models are trained using the Adam optimization algorithm, as the training process of VE-SDE proposed by Song *et al*. [29]. It should be emphasized that a fixed mask configuration is employed, which ensures methodological consistency between the training and inference stages. Specifically, an initial learning rate of $1\times10^{-4}$ is set, with a warm-up phase consisting of 5000 steps and gradient clipping established at 1. The weights are initialized using the Kaiming initialization method. Model checkpoints are saved every 10,000 steps, and with the current dataset size, the performance typically converges to its optimum around 300,000 training steps. The implementation code is composed in Python, utilizing the operator discretization library (ODL) [35] and PyTorch. Computational tasks are carried out on a workstation equipped with a GPU (NVIDIA GTX 1080Ti-11GB). During the inference phase, the number of external loop iterations is set to 1450. In the internal prediction and correction processes, the decayed Langevin dynamics are employed to better sample and explore the solution space at each execution step. To quantitatively evaluate and compare the reconstruction results, three standard metrics are employed, including peak signal-to-noise ratio (PSNR), structural similarity index (SSIM) [36], and mean squared error (MSE) [37]. Higher PSNR and SSIM values or lower MSE values are indicative of better reconstruction quality. For those interested in exploring our method, the source code of MSDiff and some representative results are available at: https://github.com/yqx7150/MSDiff.

### *C. Fan-beam CT Reconstruction*

In this section, fan-beam ultra-sparse view CT reconstruction is applied using the AAPM challenge data, CIRS phantom data, and preclinical mouse dataset, with 30, 48, and 60 projection views employed for comparison. Furthermore, to validate the performance of MSDiff, it is compared with other reconstruction methods, including FBP [38], U-Net [7], FBPConvNet [39], and GMSD [17]. FBP is a classic technique widely used in CT image reconstruction, but its performance in sparse views with low sampling rate is much worse than that in full sampling. U-Net and FBPConvNet, both supervised deep reconstruction methods, are utilized wherein U-Net is recognized as a neural network architecture for interpolating single-slice images from sparse sampling, followed by CT image reconstruction using the FBP algorithm. FBPConvNet is a combination of FBP with deep learning techniques to enhance CT reconstruction effects, while GMSD is known as an unsupervised deep learning reconstruction method capable of generating high-quality images. Tables I-II present the quantitative results reconstructed from 30, 48, and 60 views, where the best PSNR, SSIM, and MSE values for images reconstructed from different projection views are highlighted in bold. Visual results are shown in Figs. 5-8.

***AAPM Challenge Data Study:*** Table I. demonstrates that the proposed MSDiff method surpasses other reconstruction techniques across these evaluation metrics. Particularly, MSDiff exhibits significant advantages in improving image quality compared to existing technologies. MSDiff is prominently recognized for its enhancement of PSNR compared to the FBP technique. Furthermore, the visual performance of MSDiff aligns well with its outstanding quantitative evaluation metrics. As demonstrated in Figs. 5 and 6, with the increase in the number of projection views, an improvement in image quality reconstructed by MSDiff is observed, whereas images reconstructed by FBP and U-Net methods are seen to possess lower quality due to the blurriness of key details and structures. Specifically, the shortcomings in recovering complex details by FBPConvNet and the edge blurriness by GMSD are elaborated. In contrast, images produced by MSDiff are not only rich in detail but also accurately depict contour information, showcasing its superior capabilities in detail recovery and structure representation.

As shown in Fig. 6, a leading position in providing finesse in detail and structural information is consistently maintained by MSDiff, exhibiting superior visual effects compared to other methods. The validates not only does MSDiff outperform other methods in quantitative evaluation, but also holds a significant advantage in visual effects. Compared to GMSD, which primarily focuses on learning the overall characteristics of sine curves, further refinement in detail processing is achieved by MSDiff by integrating SDM, resulting in more accurate and comprehensive reconstruction. This approach enables MSDiff to capture richer feature information, thereby reconstructing complete and clear image contours.

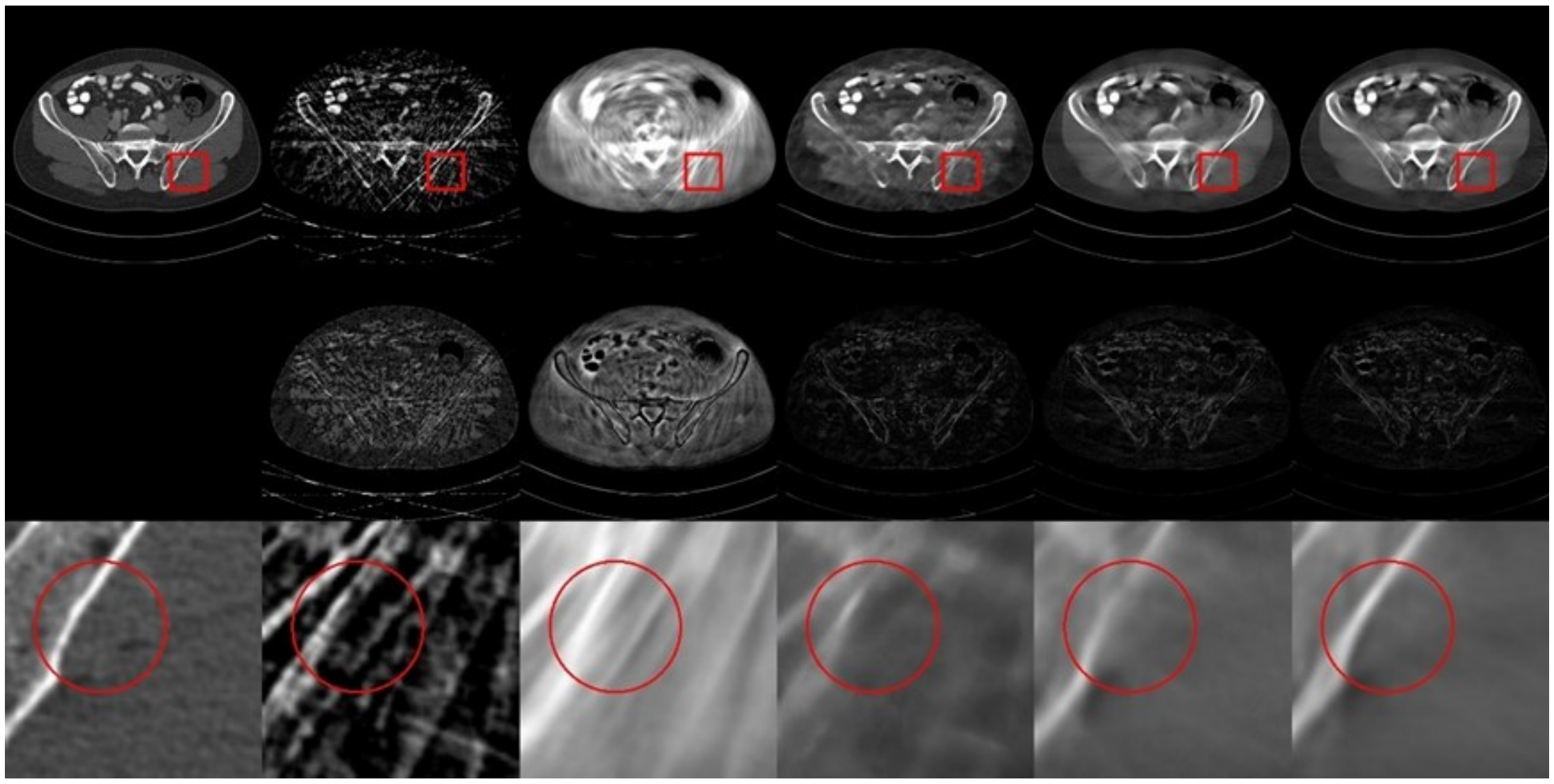

(a) (b) (c) (d) (e) (f)

**Fig. 5.** Reconstruction images with 30 views using different methods from AAPM Challenge Data. (a) The GT versus the images reconstructed by (b) FBP, (c) U-Net, (d) FBPConvNet, (e) GMSD, and (f) MSDiff. Display windows are all set to [-80, 80] HU. The second row shows the difference images.

Table I

RECONSTRUCTION PSNR/SSIM/MSE($10^{-3}$) OF AAPM CHALLENGE DATA USING DIFFERENT METHODS AT 10, 20 AND 30 VIEWS.

| Views | FBP | U-Net | FBPConvNet | GMSD | MSDiff |
|---|---|---|---|---|---|
| 30 | 19.52/0.4636/11.5 | 20.23/0.8244/10.3 | 29.68/0.9197/1.1 | 31.78/0.9321/0.7 | **32.43/0.9500/0.61** |
| 48 | 22.29/0.5504/6.0 | 24.34/0.9037/4.5 | 34.11/0.9437/0.39 | 34.65/0.9603/0.41 | **35.54/0.9675/0.31** |
| 60 | 23.45/0.5964/4.6 | 26.45/0.9279/2.9 | 35.38/0.9641/0.33 | 36.10/0.9700/0.28 | **36.49/0.9735/0.25** |

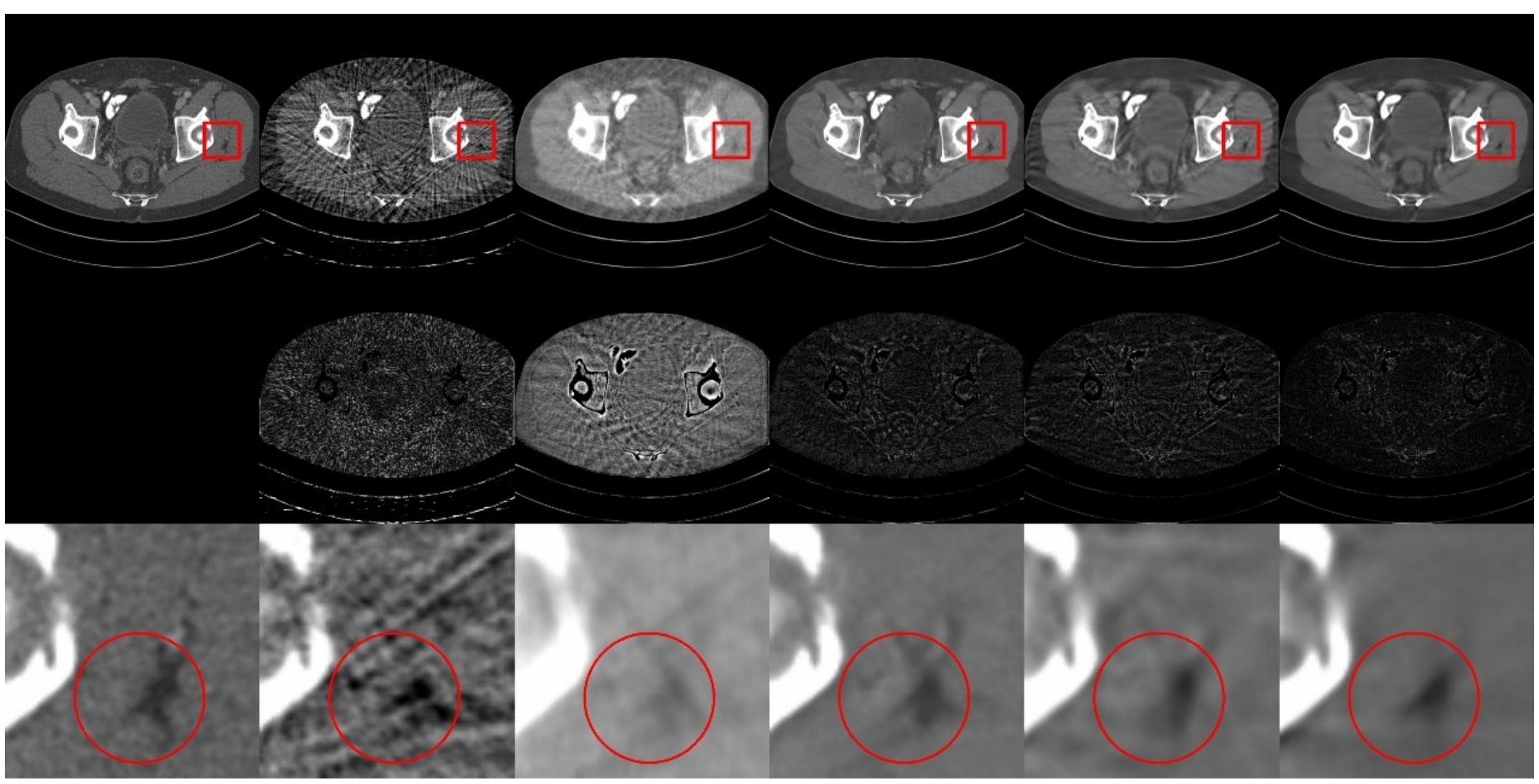

(a) (b) (c) (d) (e) (f)

**Fig. 6.** Reconstruction images with 60 views using different methods from AAPM Challenge Data. (a) The GT versus the images reconstructed by (b) FBP, (c) U-Net, (d) FBPConvNet, (e) GMSD, and (f) MSDiff. Display windows are all set to [-170, 250] HU. The second row shows the difference images.

## *D. Generalization Test*

***CIRS Phantom Data Study:*** To validate the generalization performance, the model trained on the AAPM dataset is used to test the CIRS Phantom data. Table II presents the quantitative results obtained from the CIRS Phantom data. Notably, the MSDiff method surpasses other comparative methods in terms of the highest quantitative metrics. MSDiff exhibits higher structural similarity on the CIRS Phantom dataset. Specifically, higher accuracy, stronger detail preservation capability, and clearer image reconstruction ability are consistently maintained by MSDiff. This robust performance highlights the effectiveness of MSDiff in surpassing GMSD.

Table II
RECONSTRUCTION PSNR/SSIM/MSE($10^{-3}$) OF CIRS PHANTOM DATA USING DIFFERENT METHODS AT 10, 20, AND 30 VIEWS.

| Views | FBP | U-Net | FBPConvNet | GMSD | MSDiff |
|---|---|---|---|---|---|
| 10 | 12.02/0.2846/62.8 | 19.99/0.7293/10.0 | 20.67/0.7566/8.7 | 20.83/**0.7877**/8.3 | **21.72**/0.7745/**6.8** |
| 20 | 12.71/0.3312/53.7 | 21.75/0.7995/6.7 | 21.89/0.7962/6.5 | 23.29/0.8616/4.7 | **24.86/0.8808/3.3** |
| 30 | 12.77/0.4159/52.8 | 25.76/0.8786/2.7 | 24.43/0.8193/3.7 | 26.67/0.9053/2.2 | **26.82/0.9243/2.1** |

Moreover, exceptional visual performance is exhibited by MSDiff. In Figs. 7, the enlarged regions of interest are highlighted with red rectangles. Under 20 and 30 viewing angles, significant streak artifacts, compared to FBP, are reduced by unsupervised U-Net and FBPConvNet (combining FBP and deep learning). Conversely, blurred contours and unclear details are exhibited, resulting in poor visual outcomes. Additionally, while a notable improvement in reducing artifacts is shown by GMSD, some texture information is also sacrificed. In contrast, outstanding reconstruction capabilities are demonstrated by the MSDiff method, effectively reducing artifacts in ultra-sparse view CT and restoring image detail information. Compared to other similar reconstruction methods, textures and details are excellently preserved by MSDiff, offering optimal visual effects.

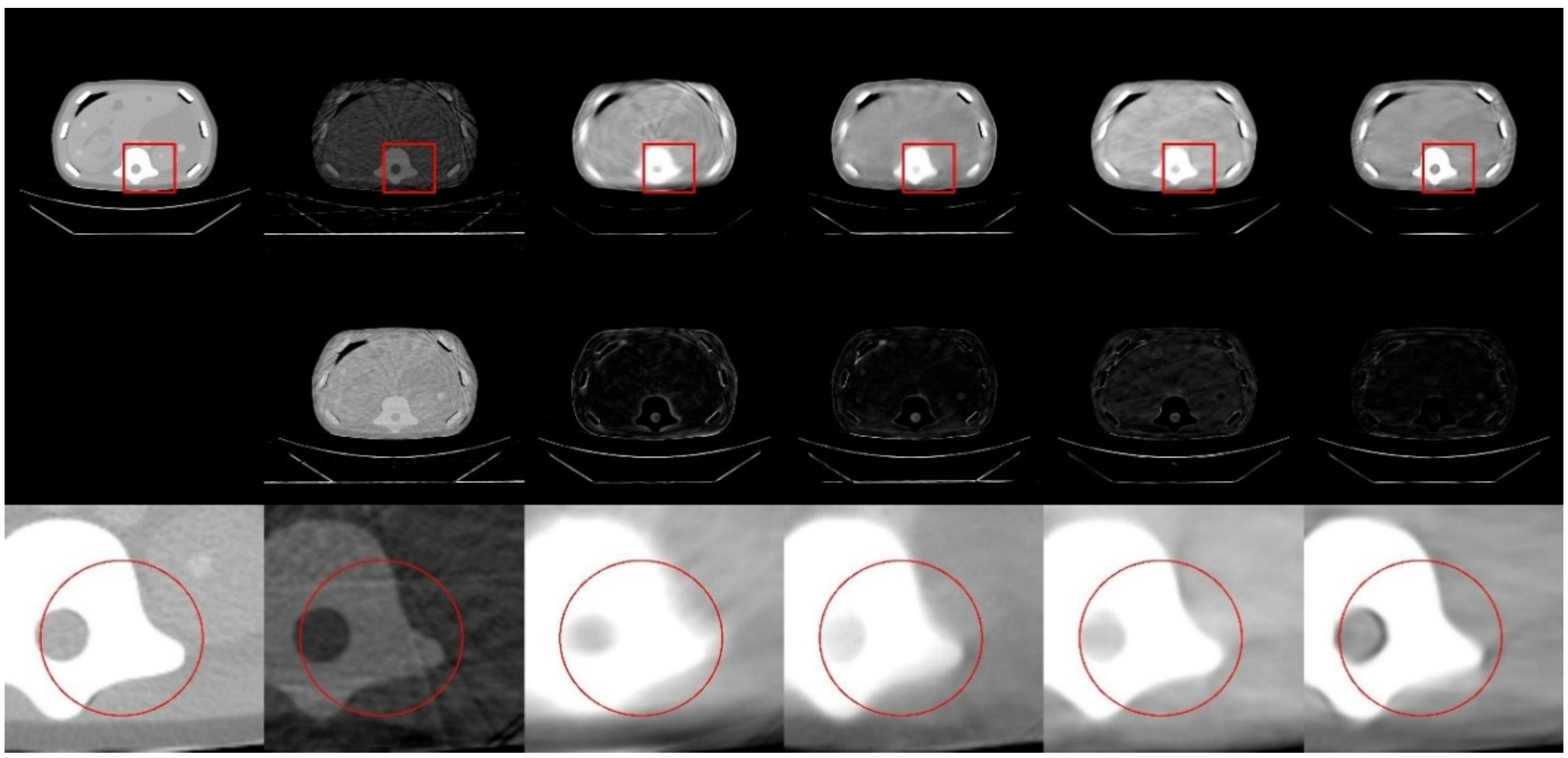

**Fig. 7.** Reconstruction images with 30 views from CIRS Phantom Data. (a) The GT versus the images reconstructed by (b) FBP, (c) U-Net, (d) FBPConvNet, (e) GMSD, and (f) MSDiff. Display windows are all set to [-80, 80] HU. The second row shows the difference images, and the third row displays the extracted ROI.

***Preclinical Mouse Reconstruction Results:*** To evaluate efficiency on real data, the network within MSDiff is trained on AAPM challenge data and tested on preclinical mouse data. Fig. 8 visualizes and presents the reconstructed images using different methods. The reference reconstruction is originated from full-projection data. Notably, Fig. 8 vividly displays the streaking artifacts that are present in FBP reconstruction. Upon examining the extracted regions of interest (ROIs), it is observed that both U-Net and FBPConvNet tend to produce blurred structures, which affect the clarity of details within the ROIs. The results in image structures deviating from the original images, indicating a significant deficiency in preserving complex structures. In contrast, the reconstructed results of GMSD show the general image structure. Compared to GMSD, approximate edges are displayed by MSDiff, with image structures being closer to the GT. Table III provides a detailed quantitative analysis of the clinical mouse data, highlighting the improvements of MSDiff over GMSD across various metrics.

Table III
RECONSTRUCTION PSNR/SSIM/MSE($10^{-3}$) OF PRECLINICAL MOUSE DATA USING DIFFERENT METHODS AT 50 VIEWS.

| Methods | PSNR | SSIM | MSE($10^{-3}$) |
|---|---|---|---|
| FBP | 24.25 | 0.5105 | 3.76 |
| U-Net | 28.14 | 0.8474 | 1.54 |
| FBPConvNet | 26.23 | 0.6110 | 2.38 |
| GMSD | 33.28 | 0.9200 | 0.47 |
| MSDiff | **34.04** | **0.9204** | **0.40** |

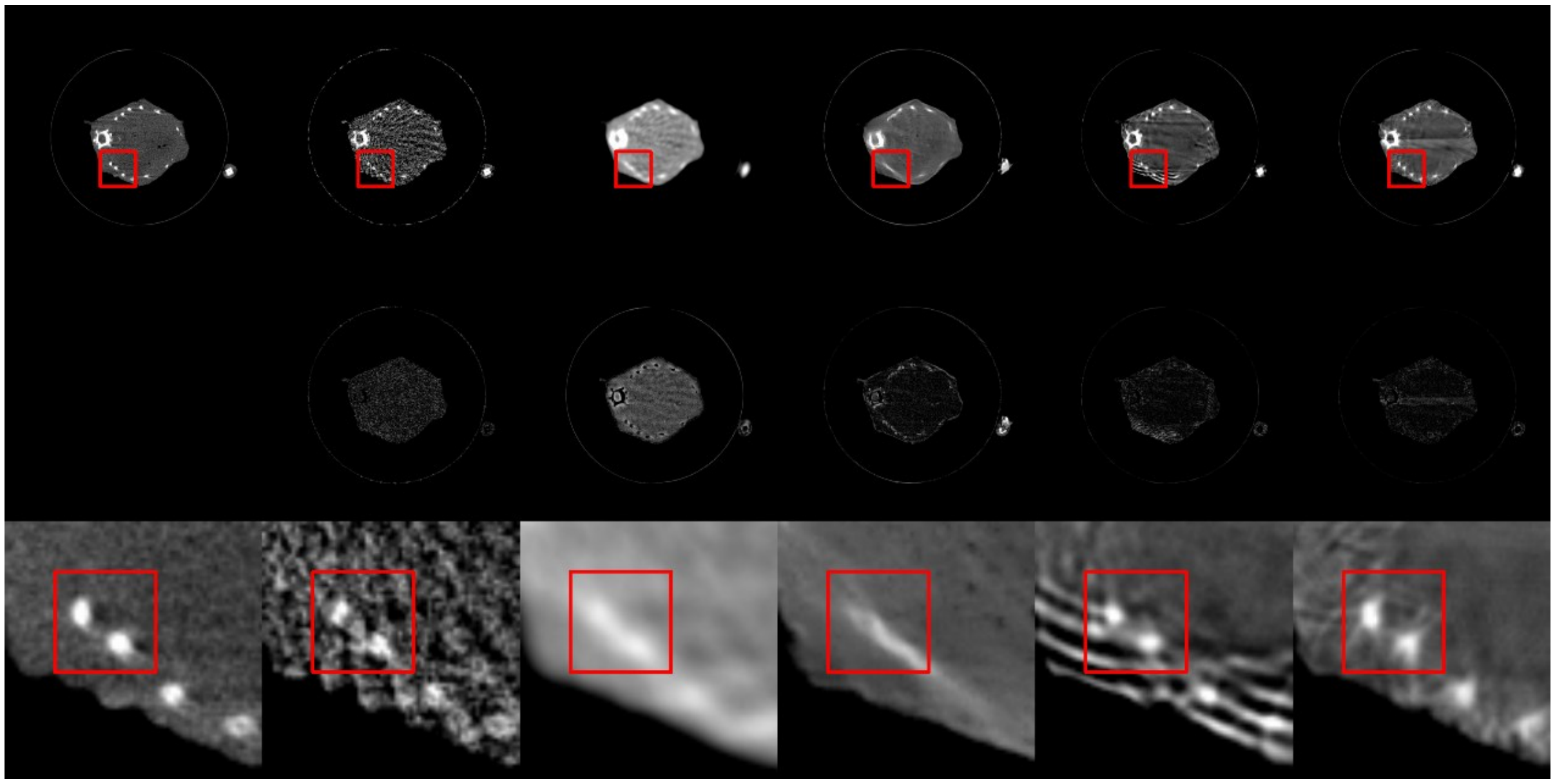

(a) (b) (c) (d) (e) (f)

**Fig. 8.** Reconstruction images with 50 views from Preclinical Mouse Data. (a) The GT versus the images reconstructed by (b) FBP, (c) U-Net, (d) FBPConvNet, (e) GMSD, and (f) MSDiff. Display windows are all set to [-250, 220] HU. The second row shows the difference images.

### *E. Profile Lines Analysis*

In this section, the profile lines of reconstructed results obtained by different methods are compared to evaluate their edge-preserving performance. As shown in Fig. 9, the profile lines generated by FBP, U-Net, and FBPConvNet do not match well with the profile lines of the ground truth. Meanwhile, the profile lines produced by GMSD and MSDiff are found to be closer to the ground truth, demonstrating the strong profile-preserving capability of diffusion models. In a comprehensive comparison, the most accurate profile lines are generated by MSDiff. Additionally, it is observed that the image profile lines produced in areas with significant gray-scale changes are noticeably different. The limited availability of anatomical information on CT images could be a possible reason for this discrepancy. Despite these challenges, Fig.9 shows a significant agreement between the image profile lines generated by this method and the ground truth data, indicating the robust profile-preserving capability of MSDiff.

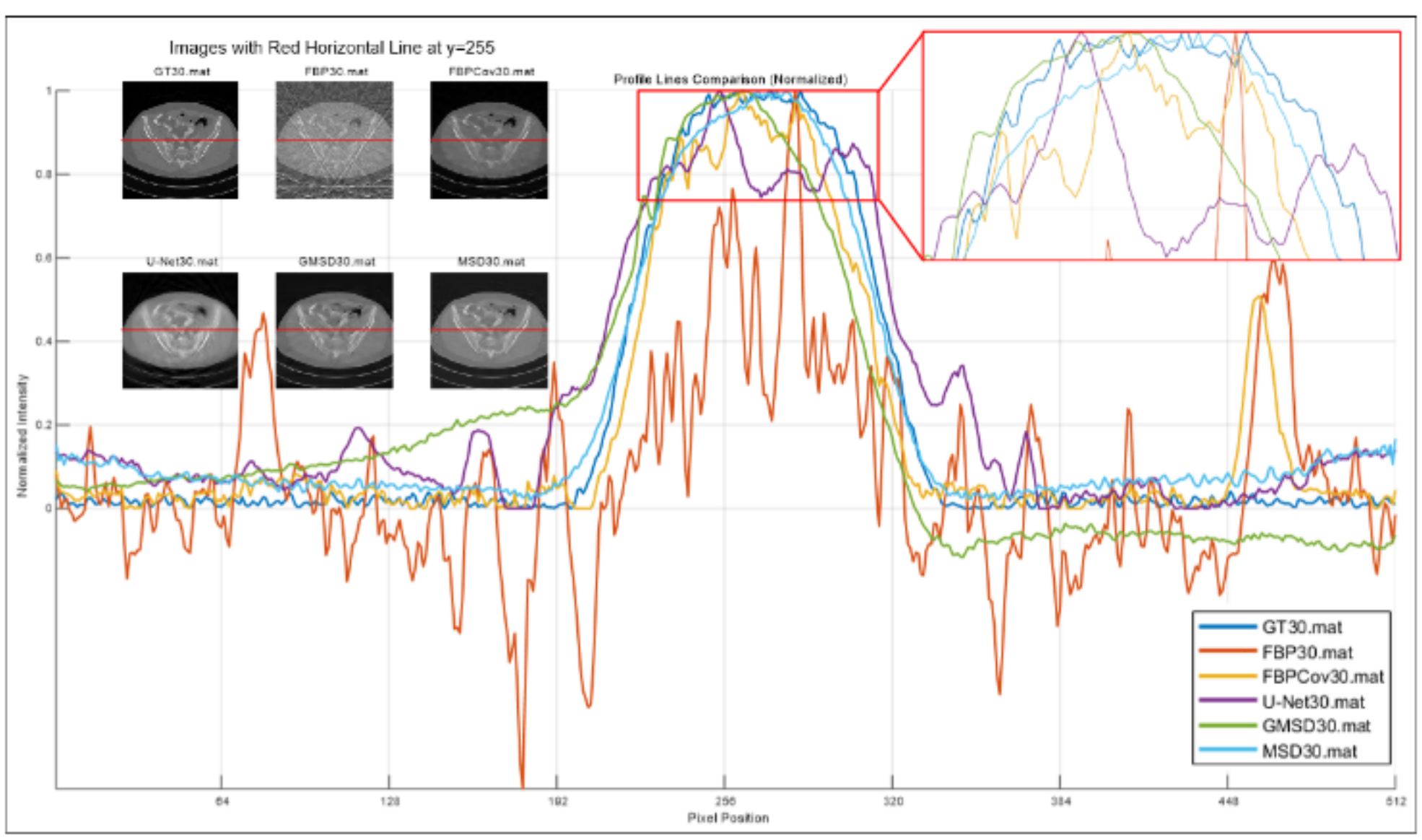

**Fig. 9.** Comparison of CT image profiles between ground truth and other methods on abdomen images. Red lines represent the location where the line profile is measured.

### *F. Ablation Study*

This work introduces a MSDiff strategy that integrates the strengths of both FDM and SDM, focusing on capturing global coherence and precise local details of images. Through detailed ablation experiments, the specific impacts of FDM and SDM on image reconstruction outcomes are explored. It should be noted that all modules are configured to have a comparable number of learnable parameters, ensuring a fair evaluation of their respective contributions. As demonstrated in Table IV, a significant advantage in enhancing image reconstruction performance through the use of the multi-scale diffusion model is shown. The combination of FDM and SDM not only preserves the individual strengths of each model but also leads to noticeable improvements in the quality of reconstructions.

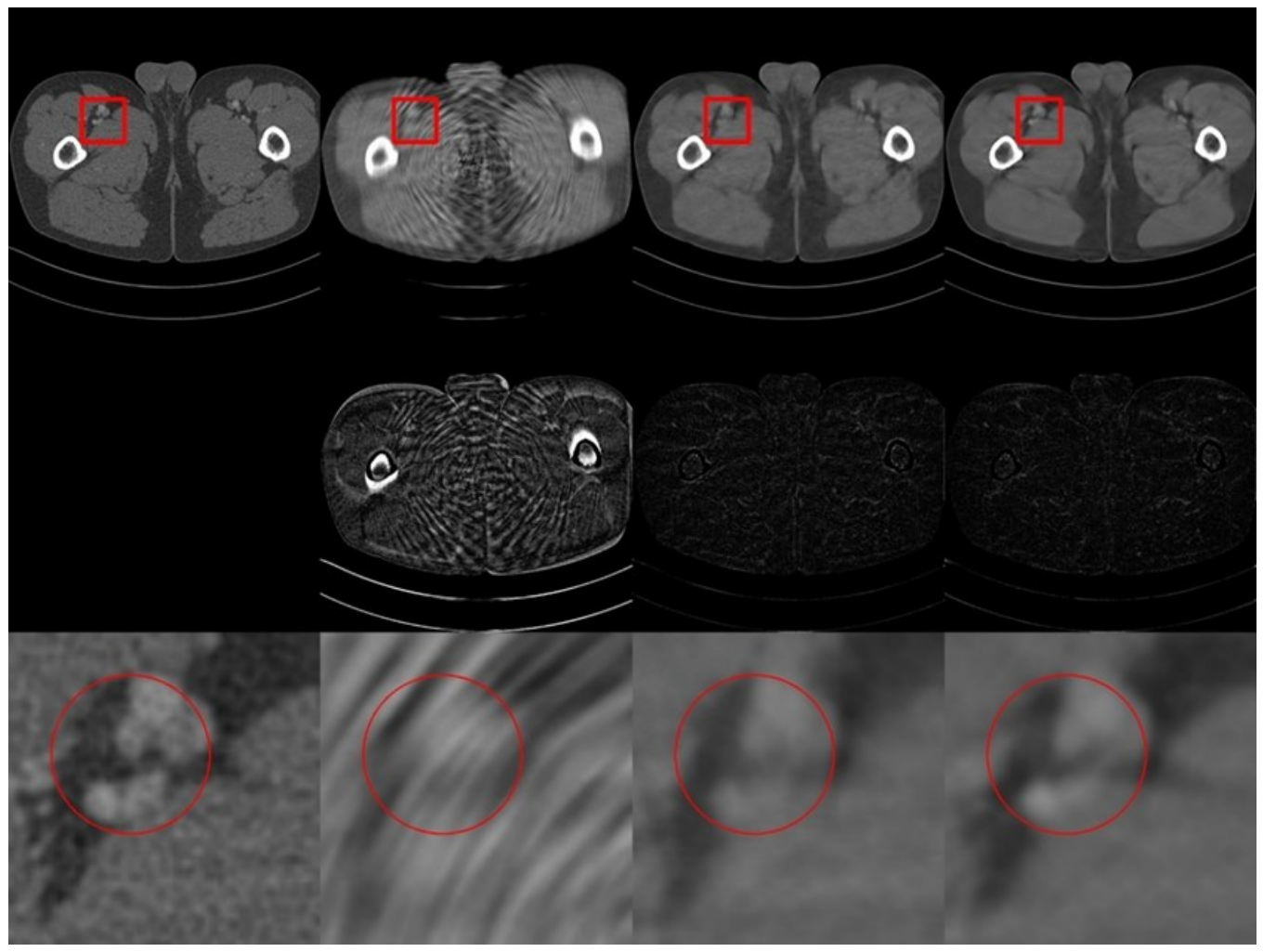

**Fig. 10.** Reconstructed images from 30 different perspectives obtained using different methods: (a) GT, (b) MSDiff without the FDM, (c) MSDiff without the SDM, and (d) fully implemented MSDiff. Display windows are all set to [-180, 250] HU.

To further elucidate the effects brought by the integration of models, qualitative analysis results in Fig. 10 reveal the clear distinction between employing and not employing the SDM module within the MSDiff model. Versions without the integrated FDM module display disorganized image structures and a lack of detailed features. Conversely, the MSDiff model lacking the SDM component is shown to perform poorly in reproducing the global outlines of images, primarily due to the absence of a refinement process for enhancing fine structures and textures. This shortcoming becomes particularly evident in differential images, highlighting the challenges of restoring the complete image framework relying solely on local information. Through the effective integration of FDM and SDM, the MSDiff model not only optimizes the details and global information in image reconstructions but also demonstrates its strong capability for comprehensive reconstruction using complementary strategies.

Table IV
RECONSTRUCTION RESULTS ON AAPM CHALLENGE DATA.

| Methods | Views | PSNR | SSIM | MSE($10^{-3}$) |
|---|---|---|---|---|
| FDM | 10 | 21.66 | 0.7013 | 6.2 |
| | 20 | 27.03 | 0.8597 | 2.1 |
| | 30 | 30.27 | 0.9238 | 1.0 |
| SDM | 10 | 17.72 | 0.6771 | 19.0 |
| | 20 | 20.12 | 0.7467 | 10.8 |
| | 30 | 21.57 | 0.7716 | 7.5 |
| MSDiff | **10** | **23.96** | **0.8092** | **4.5** |
| | **20** | **29.02** | **0.9028** | **1.3** |
| | **30** | **32.48** | **0.9504** | **0.6** |

## V. CONCLUSIONS

Despite the significant advancements achieved by deep learning techniques in the field of CT image reconstruction, the challenge of reconstructing ultra-sparse CT views persisted. Addressing this issue, a novel approach for ultra-sparse view CT image reconstruction, termed as MSDiff, was introduced in this study. This involved simultaneously training score networks under both fully sampled and under-sampled conditions, ingeniously capturing image priors across different scales. The precision of image reconstruction was substantially enhanced through an iterative update mechanism that alternated between the utilization of both networks for image prediction and fidelity enhancement. Moreover, the performance of the model was further refined with the incorporation of diverse sparse model combinations and the introduction of perception-based loss functions. Tested across three public datasets, the MSDiff approach was demonstrated to surpass existing single-network models in both quantitative assessments and visual outcomes. Overall, this research validated the effectiveness of using an iterative approach with multiple diffusion models for medical image reconstruction, presenting a feasible pathway for ultra-sparse reconstruction techniques.